\documentclass[prd,amsmath,amssymb,floatfix,superscriptaddress,nofootinbib,twocolumn]{revtex4-1}

\def\be{\begin{equation}}
\def\ee{\end{equation}}
\def\bea{\begin{eqnarray}}
\def\eea{\end{eqnarray}}

\newcommand{\vrr}{\vec{r}}

\newcommand{\vs}{\nonumber\\}

\newcommand{\vk}{\vec{k}}
\newcommand{\ec}[1]{Eq.~(\ref{eq:#1})}

\newcommand{\eql}[1]{\label{eq:#1}}
\newcommand{\rf}[1]{\ref{fig:#1}}
\newcommand{\sfig}[2]{
\includegraphics[width=#2]{#1}
        }

\newcommand{\Sfig}[2]{
   \begin{figure}[thbp]
   \begin{center}
    \sfig{#1.pdf}{\columnwidth}
    \caption{{\small #2}}
    \label{fig:#1}
     \end{center}
   \end{figure}
}

\usepackage{calligra}
\DeclareMathAlphabet{\mathcalligra}{T1}{calligra}{m}{n}
\DeclareMathAlphabet{\mathpzc}{OT1}{pzc}{m}{it}
\usepackage[utf8]{inputenc}
\usepackage{graphicx}
\usepackage{amssymb}
\usepackage{simplewick}
\usepackage{amsmath}
\usepackage{bm}
\usepackage{color}
\usepackage{enumitem}
\usepackage[linktocpage=true]{hyperref} 
\hypersetup{
    colorlinks=true,       
    linkcolor=red,         
    citecolor=blue,        
    filecolor=magenta,      
    urlcolor=blue           
}
\usepackage[all]{hypcap} 
\definecolor{darkgreen}{cmyk}{0.85,0.1,1.00,0} 
\definecolor{darkorange}{rgb}{1.0,0.2,0.0}

\begin{document}
\title{New Probes of Large Scale Structure}
\author{\large Peikai Li}
\affiliation{Department of Physics, Carnegie Mellon University, Pittsburgh, PA 15213, USA}
\affiliation{McWilliams Center for Cosmology, Carnegie Mellon University, Pittsburgh, PA 15213, USA}
\affiliation{NSF AI Planning Institute, Carnegie Mellon University, Pittsburgh, PA 15213, USA}
\author{\large Rupert A. C. Croft}
\affiliation{Department of Physics, Carnegie Mellon University, Pittsburgh, PA 15213, USA}
\affiliation{McWilliams Center for Cosmology, Carnegie Mellon University, Pittsburgh, PA 15213, USA}
\affiliation{NSF AI Planning Institute, Carnegie Mellon University, Pittsburgh, PA 15213, USA}
\author{\large Scott Dodelson}
\affiliation{Department of Physics, Carnegie Mellon University, Pittsburgh, PA 15213, USA}
\affiliation{McWilliams Center for Cosmology, Carnegie Mellon University, Pittsburgh, PA 15213, USA}
\affiliation{NSF AI Planning Institute, Carnegie Mellon University, Pittsburgh, PA 15213, USA}
\date{\today}

\begin{abstract}
\noindent This is the second paper in a series where we propose a method of indirectly measuring large scale structure using information from small scale perturbations. The idea is to build a quadratic estimator from small scale modes that provides a map of structure on large scales.
We demonstrated in the first paper that the quadratic estimator works well on a dark-matter-only N-body simulation at a snapshot of $z=0$. Here we generalize the theory to the case of a light cone halo catalog with a non-cubic region taken into consideration. We successfully apply the generalized version of the quadratic estimator to the light cone halo catalog based on an N-body simulation of volume $\sim15.03\,(h^{-1}\,\rm Gpc)^3$. The most distant point in the light cone is at a redshift of $1.42$, indicating the applicability of our method to next generation of galaxy surveys.
\end{abstract}
\maketitle

\section{Introduction} \label{sec1}
\noindent Directly measuring the distribution of matter on large
scales is extremely difficult due to observational and astrophysical
limitations.  For example,
\cite{Modi:2019hydr} point out how large spatial scales in neutral
hydrogen surveys are
completely contaminated by astrophysical foregrounds. Attempts
to use small scale perturbations to get around these limitations and
infer large scale information has been frequently discussed in recent
years, by \cite{Modi:2019hydr}, and others:
\cite{Baldauf:2011fer}\cite{Jeong:2012foss}\cite{Li:2014ssc}\cite{Zhu:2016tidal}\cite{Barreira:2017res}. In
our first work \cite{Li:2020fir}, we proposed a method for indirectly
measuring large scale structure using the small scale density
contrast. Physically, long- and short-wavelength modes are correlated
because small scale modes grow differently depending on the large
scale structure they reside in. This phenomenon leaves a signature in
Fourier space: the two-point statistics of short-wavelength matter
density modes will have non-zero off-diagonal terms proportional to
long-wavelength modes. This is our starting point for constructing the
quadratic estimator for long-wavelength modes. We tested the power of
this quadratic estimator using a dark-matter-only catalog from an
N-body simulation in our first paper. In the present work, we generalize
Ref.~\cite{Li:2020fir} to account for two main effects that must be
accounted for before applying the techniques to upcoming
surveys, e.g., \cite{LSST:2012ls}\cite{Wfirst:2012jg}\cite{DESI:2019ds}: (i)
we observe galaxies, not the dark matter field; (ii) we observe a
non-cubic light cone rather than a single redshift snapshot. After dealing
with these, we should
be able to apply our
method to real surveys in the near future.

We first need to account for galaxy bias
\cite{Kravtsov:1999hb}\cite{Desjacques:2018rev}. Galaxy bias is a term
relating the galaxy number density contrast to the matter density
contrast \cite{Gil-Marin:2014sta}\cite{Gil-Marin:2016wya}. We use a model of second order bias, as done in recent treatments of galaxy
surveys. Meanwhile, analytically the generalization to even higher
order biases is straightforward. We adopt the most commonly used
second-order galaxy bias model and assume all the bias parameters to
be constants even while we are considering a large volume across a wide
redshift range.

Observationally a galaxy catalog will be in the form of a light cone
\cite{Carroll:1997gr} instead of a single redshift snapshot.
The typical treatment is
to cut a light cone into several thin redshift bins
\cite{Chuang:2016uuz} and analyze the properties within each
bin. Doing this, though, leads to loss of information on the long-wavelength
modes along the line of sight. Thus in this paper we propose a method of
considering all the galaxies in a light cone together, using the
well-known Feldman-Kaiser-Peacock (FKP) estimator \cite{Feldman:1993ky} to
account for the evolution of the galaxy number density. Using an
octant volume (the technique can be applied to a even
more generalized shape), we
test the quadratic estimator for long-wavelength modes using
information from non-zero off-diagonal terms as in
\cite{Li:2020fir}. It should be  noticed
that the FKP description corresponds to
the monopole part of the estimator in redshift space (e.g. the
Yamamoto estimator \cite{Yamamoto:2005dz}\cite{Bianchi:2015oia}).
Because of this,
our formalism will be able to reconstruct the large scale monopole power
spectrum which is the main goal when studying the large scale matter
distribution of the 3D universe.


We begin with a brief review of the formalism developed in
Ref.~\cite{Li:2020fir}, then present our treatment of the galaxy
number density contrast in a light cone and build the quadratic
estimator. Finally we apply the estimator to light-cone halo simulations and extract the large scale modes accounting for
these effects.
\clearpage


\section{Review of Quadratic Estimator} \label{sec2}

\noindent We first review the construction of a quadratic estimator of
a dark-matter-only catalog \cite{Li:2020fir} before moving to a halo
catalog. We start from the perturbative expansion of the matter density
contrast in Fourier space up to second order
\cite{Jain:1994sop}\cite{Bernardeau:2002rev}: \bea &&\delta_{\rm
  m}(\vk;z)=\delta^{(1)}_{\rm m}(\vk;z)+\delta^{(2)}_{\rm m}(\vk;z)\vs
&=&\delta^{(1)}_{\rm m}(\vk;z)\vs &+&\int
\frac{d^{3}\vec{k}'}{(2\pi)^3}\delta^{(1)}_{m}(\vec{k}';z)\delta^{(1)}_{m}(\vec{k}-\vec{k}';z){F}_{2}(\vec{k}',\vec{k}-\vec{k}')
\eql{tayl} \eea where ``m" stands for matter, the superscript
$i=1,2,\cdots$ corresponds to the $i$-th order term of the expansion,
and $\delta_{\rm m}(\vk;z)$ is the full Fourier space matter density
contrast in a snapshot at redshift $z$. The
kernel $F_{2}$ is a function particularly insensitive to the choice of
cosmological parameters in a dark-energy-dominated universe
\cite{Takahashi:2008to}: \be
F_{2}(\vk_1,\vk_2)=\frac{5}{7}+\frac{2}{7}\frac{(\vk_1\cdot
  \vk_2)^2}{k_1^2 k_2^2}+\frac{\vk_1\cdot
  \vk_2}{2k_1k_2}\bigg[\frac{k_1}{k_2}+\frac{k_2}{k_1}\bigg].\eql{f2}
\ee Thus, $\delta_{\rm m}^{(1)}$ is the linear density contrast, and
the second order term $\delta_{\rm m}^{(2)}$ can be written as a
convolution-like integral using the first order term.

When evaluating the two-point function of the full density contrast,
cross-terms appear. For example, $\langle \delta_{\rm m}^{(1)}(\vk;z)
\delta_{\rm m}^{(2)}(\vk';z) \rangle$ is proportional to
$\delta^{(1)}_{\rm m}(\vk+\vk';z)$ if both $\vk$ and $\vk'$ correspond
to short wavelengths but their sum is small (long
wavelength). Explicitly, keeping terms up to second order, \be \langle
{\delta}_{\rm m}(\vec{k}_s;z){\delta}_{\rm m}(\vec{k}_s';z) \rangle
=f(\vec{k}_s,\vec{k}_s';z){\delta}^{(1)}_{\rm
  m}(\vec{k}_l;z). \eql{2pt} \ee Here $\vk_s$ and $\vk_s'$ are two
short-wavelength modes and $\vk_l$ is a long-wavelength mode
($\vk_s,\vk_s' \gg \vk_l$). They satisfy the squeezed-limit condition,
$\vk_s+\vk_s'=\vk_l$, and $f$ is given by: \bea
f(\vec{k}_s,\vec{k}_s';z)&=&2F_2(-\vec{k}_s,\vec{k}_s+\vec{k}_s')P_{\rm
  m}^{(1)}(k_s;z)\vs &+&2F_2(-\vec{k}_s',\vec{k}_s+\vec{k}_s')P_{\rm
  m}^{(1)}(k_s';z) .\eql{deff} \eea Here $P_{\rm m}^{(1)}$ is the
linear matter power spectrum. \ec{2pt} indicates that we can estimate
the long-wavelength modes using small scale information with the
following minimum variance quadratic estimator:
\begin{eqnarray}
\hat{\delta}^{(1)}_{\rm m}(\vec{k}_l;z)=A(\vec{k}_l;z)\int \frac{d^3 \vec{k}_s}{(2\pi)^3} g(\vec{k}_s,\vec{k}_s';z){\delta}_{\rm m}(\vec{k}_s;z){\delta}_{\rm m}(\vec{k}_s';z)\vs \eql{quadest}
\end{eqnarray} 
with $\vk_s'=\vk_l-\vk_s$. The normalization factor $A$ is defined by requiring that $\langle \hat{\delta}_{\rm m}^{(1)}(\vec{k}_l;z) \rangle={\delta}_{\rm m}^{(1)}(\vec{k}_l;z)$, and the weighting function $g$ is obtained by minimizing the noise. They can be expressed as:
\begin{eqnarray}
A(\vec{k}_l;z)&=&\bigg[\int \frac{d^3 \vec{k}_s}{(2\pi)^3} g(\vec{k}_s,\vec{k}_s';z)f(\vec{k}_s,\vec{k}_s';z)  \bigg]^{-1} \eql{a} \vs
g(\vec{k}_{s},\vec{k}_{s}';z)&=&\frac{f(\vec{k}_{s},\vec{k}_{s}';z)}{2P_{\rm m}(k_{s};z)P_{\rm m}(k_{s}';z)}
\end{eqnarray}
where $P_{\rm m}$ is the nonlinear matter power spectrum including shot noise. With this choice of the weighting function $g$, the noise on the estimator $N(\vk_l;z)=A(\vk_l;z)$ if non-Gaussian terms in the four-point function are neglected. Therefore, the projected detectability of a power spectrum measurement using this quadratic estimator can be written as:
\be
\frac{1}{\sigma^{2}(k_l;z)}=\frac{V k_l^2 \Delta k }{(2\pi)^2}\bigg[\frac{P_{\rm m}^{(1)}(k_l;z)}{P_{\rm m}^{(1)}(k_l;z)+A(k_l;z) }\bigg]^2 \eql{error},
\ee
where $V$ is the volume of a survey and $\Delta k$ is the width of long-wavelength mode bins. We also take advantage of the fact that $A(\vk_l;z)$ does not depend on the direction of the long mode $\vk_l$.

\section{Generalization: Bias Model and FKP estimator} \label{sec3}
Galaxy bias describes the statistical relation between dark-matter and galaxy distributions. Similar to \ec{tayl}, we use the most commonly used Eulerian
non-linear and non-local galaxy bias model np to second-order first
proposed by \cite{McDonald:2009dh}: \bea &&\delta_{\rm g}(\vec{k};z) =
b_1\delta^{(1)}_m (\vec{k};z)\nonumber \\ &+&\int
\frac{d^{3}\vec{k}'}{(2\pi)^3}\delta^{(1)}_{m}(\vec{k}';z)\delta^{(1)}_{m}(\vec{k}-\vec{k}';z)\mathpzc{F}_{2}(\vec{k}',\vec{k}-\vec{k}')
\;. \eql{hdc} \eea here ``$\rm g$" denotes galaxy, and $b_{1}$ is the
linear bias parameter relating galaxy and the matter density
contrasts. The kernel $\mathpzc{F}_2$ is given by: \be
\mathpzc{F}_2(\vk_1,\vk_2)=b_1F_{2}(\vk_1,\vk_2)+\frac{b_2}{2}+\frac{b_{s^2}}{2}S_{2}(\vk_1,\vk_2)
\ee with $S_2$ given by: \be S_{2}(\vk_1,\vk_2)=\frac{(\vk_1\cdot
  \vk_2)^2}{k_1^2k_2^2}-\frac{1}{3}\;.  \ee
Comparing the perturbative
expansion of the galaxy density contrast \ec{hdc} with that of the matter
density contrast \ec{tayl}, we see that the difference with the first
order term is an extra coefficient $b_1$.  The second order term is
also almost the same, with a simple replacement of the kernel function. This
implies that we can easily generalize to the case of a galaxy catalog
in a snapshot. In the case of a galaxy catalog in a light cone,
the Feldman-Kaiser-Peacock (FKP) estimator is usually used to construct a weighted over-density that can be used to obtain
the observed galaxy power spectrum \cite{Feldman:1993ky}: \be
F(\vec{r})\equiv I^{-1/2}w_{\rm FKP}(\vec{r})[n_{g}(\vec{r})-\alpha
  n_{s}(\vec{r})] \ee with \be I\equiv\int_{V} d^{3}\vec{r}w_{\rm
  FKP}^{2}(\vec{r})\langle n_{ g}\rangle^{2}(\vec{r})\;.  \ee Here
$n_{g}$ is the observed galaxy number density and $n_{s}$ is the
corresponding synthetic catalog (a random catalog with the same angular and
radial selection function as the observations). The constant $\alpha$ is the ratio of
the observed number density to the synthetic catalog's number
density. The FKP weight $w_{\rm FKP}(\vec{r})$ is usually defined as:
\begin{eqnarray}
w_{\rm FKP}(\vec{r})=\frac{1}{1+\langle n_{g}\rangle (\vec{r})P_{0}}
\end{eqnarray}
where $P_0$ is the typical amplitude of the observed power spectrum at the scale where the signal-to-noise of the power spectrum estimation is maximized, usually $k\sim 0.12\,h \rm Mpc^{-1}$. Note that in real surveys we will have other types of weights \cite{Gil-Marin:2014sta}\cite{Gil-Marin:2018cgo}, which can be easily included in the formalism of this section. The FKP estimator $F(\vec{r})$ is related to the observed galaxy power spectrum $P_{g,\rm obs}(\vec{k})$ by considering the following expectation value (diagonal elements) in Fourier space:
\begin{eqnarray}
\langle |F(\vec{k})|^{2}\rangle &=&\int \frac{d^{3}\vec{k}'}{(2\pi)^3}P_{g}(k';z_{\rm eff})|W(\vec{k}-\vec{k}')|^2+P_{\rm shot \; noise}\nonumber\\
&=&\langle |\delta_{g,W}(\vec{k};z_{\rm eff})|^2 \rangle+P_{\rm shot \; noise}\equiv P_{g,\rm obs}(\vec{k}),
\end{eqnarray}
where $z_{\rm eff}$ is the effective redshift of the whole light cone. The window function $W(\vk)$, the shot noise spectrum $P_{\rm shot \; noise}$ and the windowed galaxy density contrast $\delta_{g,W}(\vk;z)$ are given respectively by\footnote{An interesting thing to notice here is that in the original paper introducing the FKP estimator \cite{Feldman:1993ky} and in almost every
  succeeding work, $W(\vk)$ is defined as the complex conjugate of the quantity we use here. In these other cases, $W(\vk)$ only appears in the form of $|W(\vk)|^2$, and  so this does not make a difference. However in our current work, we demonstrate using simulations that  $e^{-i\vec{k}\cdot \vec{r}}$ should appear instead of $e^{+i\vec{k}\cdot \vec{r}}$ in the integrand, the same as for the definition of the Fourier transform.}:
\begin{eqnarray}
W(\vec{k})&= &I^{-1/2}{\int_{V} d^{3}\vec{r} \langle n_{g}\rangle(\vec{r})w_{\rm FKP}(\vec{r})e^{-i\vec{k}\cdot \vec{r}}  }\eql{WinF}\\
P_{\rm shot \; noise}&=&(1+\alpha)I^{-1}{\int_{V}d^{3}\vec{r} \langle n_{g}\rangle (\vec{r}) w_{\rm FKP}^{2}(\vec{r})}\\
\delta_{g,W}(\vec{k})&\equiv& \int \frac{d^{3}\vec{k}'}{(2\pi)^3} \delta_{g}(\vec{k}')W(\vec{k}-\vec{k}')\eql{WinD}\;.
\end{eqnarray}
We want to calculate the off-diagonal term of the FKP estimator $F(\vk)$ given the fact that $F(\vk)$ is the observable from a galaxy light cone survey rather than $\delta_{g,W}$. Note that the two point functions of $n_{g}(\vec{r})-\alpha n_{s}(\vec{r})$ can be written as \cite{Feldman:1993ky}:
\begin{eqnarray}
&&\langle [n_{g}(\vec{r})-\alpha n_{s}(\vec{r})][n_{g}(\vec{r}')-\alpha n_{s}(\vec{r}')]\rangle\nonumber\\
&=&\langle n_{g}\rangle(\vec{r})\langle n_{g}\rangle(\vec{r}')\xi_{g}(\vec{r}-\vec{r}')+(1+\alpha)\langle n_{g}\rangle(\vec{r})\delta_{\rm D}(\vec{r}-\vec{r}')\vs
\end{eqnarray}
Assuming the squeezed limit $\vec{k}_s+\vec{k}'_s=\vec{k}_l$ and using the expression above, we can write the off-diagonal term as:
\begin{eqnarray}
&&\langle F(\vec{k}_s)F(\vec{k}'_s)\rangle\nonumber\\
&=&\langle \delta_{g,W}(\vec{k}_s;z_{\rm eff})\delta_{g,W}(\vec{k}'_s;z_{\rm eff}) \rangle+Q_{\rm shot \; noise}(\vec{k}_l),
\end{eqnarray}
with the``off-diagonal shot noise" $Q(\vk_l)$ given by:
\be 
Q(\vk_l)=(1+\alpha)I^{-1}{\int_{V}d^{3}\vec{r} \langle n_{g}\rangle (\vec{r}) w_{\rm FKP}^{2}(\vec{r})e^{i\vec{k}_l\cdot \vrr}}\;.
\ee 
The two point function $\langle \delta_{g,W}(\vec{k}_s;z_{\rm eff})\delta_{g,W}(\vec{k}'_s;z_{\rm eff}) \rangle$ up to second order can be simply expressed as:
\be 
\langle \delta_{g,W}\delta_{g,W}' \rangle=\langle \delta^{(1)}_{g,W}\delta'^{(1)}_{g,W} \rangle+\langle \delta^{(1)}_{g,W}\delta'^{(2)}_{g,W} \rangle+\langle \delta^{(2)}_{g,W}\delta'^{(1)}_{g,W} \rangle
\ee
by defining $\delta_{g,W}\equiv\delta_{g,W}(\vec{k}_s;z_{\rm eff})$ and $\delta_{g,W}'\equiv\delta_{g,W}(\vec{k}'_s;z_{\rm eff})$.  One major difference here is that, for a non-cubical region, the leading order term would also be non-zero unlike the cubic volume in the last section \ref{sec2}:
\bea
&&\langle \delta^{(1)}_{g,W}\delta'^{(2)}_{g,W} \rangle\vs
&=&\int \frac{d^3\vec{k}}{(2\pi)^3}\int \frac{d^3\vec{k}'}{(2\pi)^3}W(\vk-\vk_s)W(\vk'-\vk'_s)\vs
&\times & b_1^2 (2\pi)^3\delta_{\rm D} (\vk-\vk') P_m^{(1)}(k;z_{\rm eff})\vs
&=&b_1^2\int \frac{d^3\vec{k}}{(2\pi)^3}W(\vk-\vk_s)W(-\vk-\vk'_s)P_m^{(1)}(k;z_{\rm eff})
\eea 
where $\delta_{\rm D}$ is the Dirac delta function. This term would vanish since in the case of a cube, $W(\vk)$ would be close to a Dirac delta function. For a non-cubic region, though, the term is no longer zero. Note that this leading order term can be fully determined numerically.

Using the expressions from \ec{hdc} and \ec{WinD}, we can compute the second order two-point correlation of two short-wavelength modes $\delta_{g,W}(\vec{k}_s;z_{\rm eff})$ and $\delta_{g,W}(\vec{k}'_s;z_{\rm eff})$. Using $\langle \delta^{(1)}_{g,W}\delta'^{(2)}_{g,W} \rangle$ as an example, we have:
\begin{eqnarray}
&&\langle \delta^{(1)}_{g,W}\delta'^{(2)}_{g,W} \rangle\nonumber\\
&=&b_{1}\int \frac{d^{3}\vec{k}}{(2\pi)^3}\int \frac{d^{3}\vec{k}'}{(2\pi)^3}W(\vec{k}_s-\vec{k})W(\vec{k}'_s-\vec{k}')\nonumber\\
&\times & \langle \delta^{(1)}_{m}(\vec{k};z_{\rm eff})\delta^{(2)}_{g}(\vec{k}';z_{\rm eff})\rangle
\end{eqnarray}
Notice that we have computed the bracket $\langle \delta^{(1)}_{m}(\vec{k};z_{\rm eff})\delta^{(2)}_{g}(\vec{k}';z_{\rm eff})\rangle$ in Ref.~\cite{Li:2020fir}, with $F_{2}$ replaced by $\mathpzc{F}_{2}$. The result is:
\begin{eqnarray}
&&\langle \delta^{(1)}_{m}(\vec{k};z_{\rm eff})\delta^{(2)}_{g}(\vec{k}';z_{\rm eff})\rangle\vs
&=&2\mathpzc{F}_{2}(-\vec{k},\vec{k}+\vec{k}')P_{m}^{(1)}(k;z_{\rm eff})\delta_{m}^{(1)}(\vec{k}+\vec{k}';z_{\rm eff})
\end{eqnarray}
Thus we can further express the bracket as:
\begin{eqnarray}
&&\langle \delta^{(1)}_{g,W}\delta'^{(2)}_{g,W} \rangle\nonumber\\
&=&2b_{1}\int \frac{d^{3}\vec{k}}{(2\pi)^3}\int \frac{d^{3}\vec{k}'}{(2\pi)^3}W(\vec{k}_s-\vec{k})W(\vec{k}'_s-\vec{k}')\nonumber\\
&\times &\mathpzc{F}_{2}(-\vec{k},\vec{k}+\vec{k}')P_{m}^{(1)}(k;z_{\rm eff})\delta_{m}^{(1)}(\vec{k}+\vec{k}';z_{\rm eff})
\end{eqnarray}
Ideally, We would like to extract a term $\delta^{(1)}_{g,W}(\vec{k}_l;z_{\rm eff})$ from this, where:
\begin{eqnarray}
\delta^{(1)}_{g,W}(\vec{k}_l;z_{\rm eff})=b_{1}\int \frac{d^{3}\vec{k}}{(2\pi)^3}\delta_{m}^{(1)}(\vec{k};z_{\rm eff})W(\vec{k}_l-\vec{k}).
\end{eqnarray}
If one of the $W$ functions were a Dirac delta function, this would follow automatically. Here, $W$ is not a delta function, but given a large enough volume, $W(\vec{k})$ is peaked at $\vec{k}=0$ and also:
\begin{eqnarray}
\int\frac{d^{3}\vec{k}}{(2\pi)^3}W(\vec{k})= W(\vrr=0)\equiv C \eql{defC}
\end{eqnarray}
Thus we have the following approximations by first applying a redefinition of dummy variables:
\begin{eqnarray}
&&\langle \delta^{(1)}_{g,W}(\vec{k}_s;z_{\rm eff})\delta^{(2)}_{g,W}(\vec{k}'_s;z_{\rm eff})\rangle\nonumber\\
&=&2b_{1}\int \frac{d^{3}\vec{k}}{(2\pi)^3}\int \frac{d^{3}\vec{k}'}{(2\pi)^3}W(\vec{k}_s-\vec{k}+\vec{k}')W(\vec{k}'_s-\vec{k}')\nonumber\\
&\times &\mathpzc{F}_{2}(-\vec{k}+\vec{k}',\vec{k})P_{m}^{(1)}(|\vec{k}-\vec{k}'|;z_{\rm eff})\delta_{m}^{(1)}(\vec{k};z_{\rm eff})\nonumber\\
&\simeq&2C\mathpzc{F}_2(-\vec{k}_s,\vec{k}_s+\vec{k}'_s)P_{m}^{(1)}(k_s;z_{\rm eff})\nonumber\\
&\times & b_{1}\int \frac{d^{3}\vec{k}}{(2\pi)^3}\delta_{m}^{(1)}(\vec{k};z)W(\vec{k}_s+\vec{k}'_s-\vec{k})\nonumber\\
&=& 2C\mathpzc{F}_2(-\vec{k}_s,\vec{k}_s+\vec{k}'_s)P_{m}^{(1)}(k_s;z_{\rm eff})\delta^{(1)}_{g,W}(\vec{k}_l;z_{\rm eff}). \eql{extW}
\end{eqnarray}
With the calculation above, we can then recover the long-wavelength modes from the off-diagonal two-point functions of short-wavelength modes:
\begin{eqnarray}
&&\langle F(\vec{k}_s)F(\vec{k}'_s)\rangle-Q_{\rm shot \; noise}(\vk_l)\nonumber\\
&-&b_1^2\int \frac{d^{3}\vec{k}}{(2\pi)^3} W(\vec{k}-\vec{k}_s)W(-\vec{k}-\vec{k}'_s)P^{(1)}_{m}(k;z_{\rm eff})\nonumber\\
&=&\mathpzc{f}(\vec{k}_s,\vec{k}'_s;z_{\rm eff})\delta^{(1)}_{g,W}(\vec{k}_l;z_{\rm eff})
\end{eqnarray}
with
\begin{eqnarray}
\mathpzc{f}(\vec{k}_s,\vec{k}'_s;z_{\rm eff})&=&2C\mathpzc{F}_2(-\vec{k}_s,\vec{k}_s+\vec{k}'_s)P_{m}^{(1)}(k_s;z_{\rm eff})\nonumber \\
&+&2C\mathpzc{F}_2(-\vec{k}'_s,\vec{k}_s+\vec{k}'_s)P_{m}^{(1)}(k'_s;z_{\rm eff}).
\end{eqnarray}
Notice that the $\mathpzc{f}$ is almost identical to the $f$ function in section \ref{sec2}, simply with a replacement of the $F_2$ function and an extra coefficient $C$. The quadratic estimator can then be similarly formed, and is:
\begin{eqnarray}
&&\hat{\delta}^{(1)}_{g,W}(\vec{k}_l;z_{\rm eff})=\mathpzc{A}(\vec{k}_l;z_{\rm eff})\int \frac{d^{3}\vec{k}_s}{(2\pi)^3}\mathpzc{g}(\vec{k}_s,\vec{k}'_s;z_{\rm eff})\nonumber \\
&& \times \bigg[F(\vec{k}_s)F(\vec{k}'_s)-Q_{\rm shot \; noise}(\vk_l) \vs
&&- b_1^2\int \frac{d^{3}\vec{k}}{(2\pi)^3} W(\vec{k}-\vec{k}_s)W(-\vec{k}-\vec{k}'_s)P^{(1)}_{m}(k;z_{\rm eff})\bigg] \eql{qeLC}
\end{eqnarray}
with $\vec{k}'_s = \vec{k}_l-\vec{k}_s$ and $\mathpzc{g}$ being the weighting function. Notice here that the only difference is that we subtract off the non-zero leading order terms due to the non-cubical shape of the galaxy survey
volume, and these two terms can be calculated numerically. By requiring that $\langle \hat{\delta}^{(1)}_{g,W}(\vec{k}_l;z_{\rm eff})\rangle = {\delta}^{(1)}_{g,W}(\vec{k}_l;z_{\rm eff})$ we can similarly determine the normalization function $\mathpzc{A}$:
\begin{eqnarray}
\mathpzc{A}(\vec{k}_l;z_{\rm eff})=\bigg[ \int \frac{d^{3}\vec{k}_s}{(2\pi)^3}\mathpzc{g}(\vec{k}_s,\vec{k}'_s;z_{\rm eff})\mathpzc{f}(\vec{k}_s,\vec{k}'_s;z_{\rm eff}) \bigg]^{-1}.\vs
\end{eqnarray}
Similar to Ref.~\cite{Li:2020fir}, by minimizing the noise we obtain the expression for the weighting function $\mathpzc{g}$:
\begin{eqnarray}
&&\mathpzc{g}(\vec{k}_s,\vec{k}'_s;z_{\rm eff})=\frac{\mathpzc{f}(\vec{k}_s,\vec{k}'_s;z_{\rm eff})}{2P_{g,\rm obs}(\vec{k}_s)P_{g,\rm obs}(\vec{k}'_s)}\vs
&=&C\bigg[\frac{\mathpzc{F}_2(-\vec{k}_s,\vec{k}_s+\vec{k}'_s)P_{m}^{(1)}(k_s;z_{\rm eff})}{P_{g,\rm obs}(\vec{k}_s)P_{g,\rm obs}(\vec{k}'_s)}\vs
&&\, +\frac{\mathpzc{F}_2(-\vec{k}'_s,\vec{k}_s+\vec{k}'_s)P_{m}^{(1)}(k'_s;z_{\rm eff})}{P_{g,\rm obs}(\vec{k}_s)P_{g,\rm obs}(\vec{k}'_s)}\bigg].
\end{eqnarray}
Here $P_{g,\rm obs}$ is the full observed galaxy power spectrum including shot noise. With this choice of $\mathpzc{g}$ the noise term $\mathpzc{N}$ is identical to the normalization factor $\mathpzc{A}$. The projected detectability is defined as in \ec{error}:
\begin{eqnarray}
\frac{1}{\sigma(k_l;z_{\rm eff})^2}=\frac{Vk_l^2\Delta k}{(2\pi)^2}\bigg[\frac{P_{m}^{(1)}(k_l;z_{\rm eff})}{P_{m}^{(1)}(k_l;z_{\rm eff})+\mathpzc{A}(k_l;z_{\rm eff}) }\bigg]^2 .\vs\eql{errorLC}
\end{eqnarray}
Using the quadratic estimator \ec{qeLC} we can use the entirety of the small scale information from the non-cubical light cone to infer the large scale field of the windowed galaxy density contrast $\delta_{g,W}(\vrr)$.

\section{Demonstration with An N-Body Simulation} \label{sec4}
\noindent We use the MICE Grand Challenge light cone N-body simulation (MICE-GC) \cite{Fosalba:2015MI}\cite{Fosalba:2015MII}\cite{Fosalba:2013mra} to demonstrate the power of the estimator in a light cone. The catalog contains one octant of the full sky up to $z = 1.42$ (comoving distance $3062\, h^{-1}\, \rm Mpc$) without simulation box repetition, as shown in Fig.~\rf{octant}. This simulation used a flat $\Lambda$CDM model with cosmological parameters $ \Omega_{\rm m}=0.25$, $\sigma_8 = 0.8$, $n_{\rm s}=0.95$, $\Omega_{\rm b}=0.044$, $\Omega_{\Lambda}=0.75$, $h=0.7$. 

\Sfig{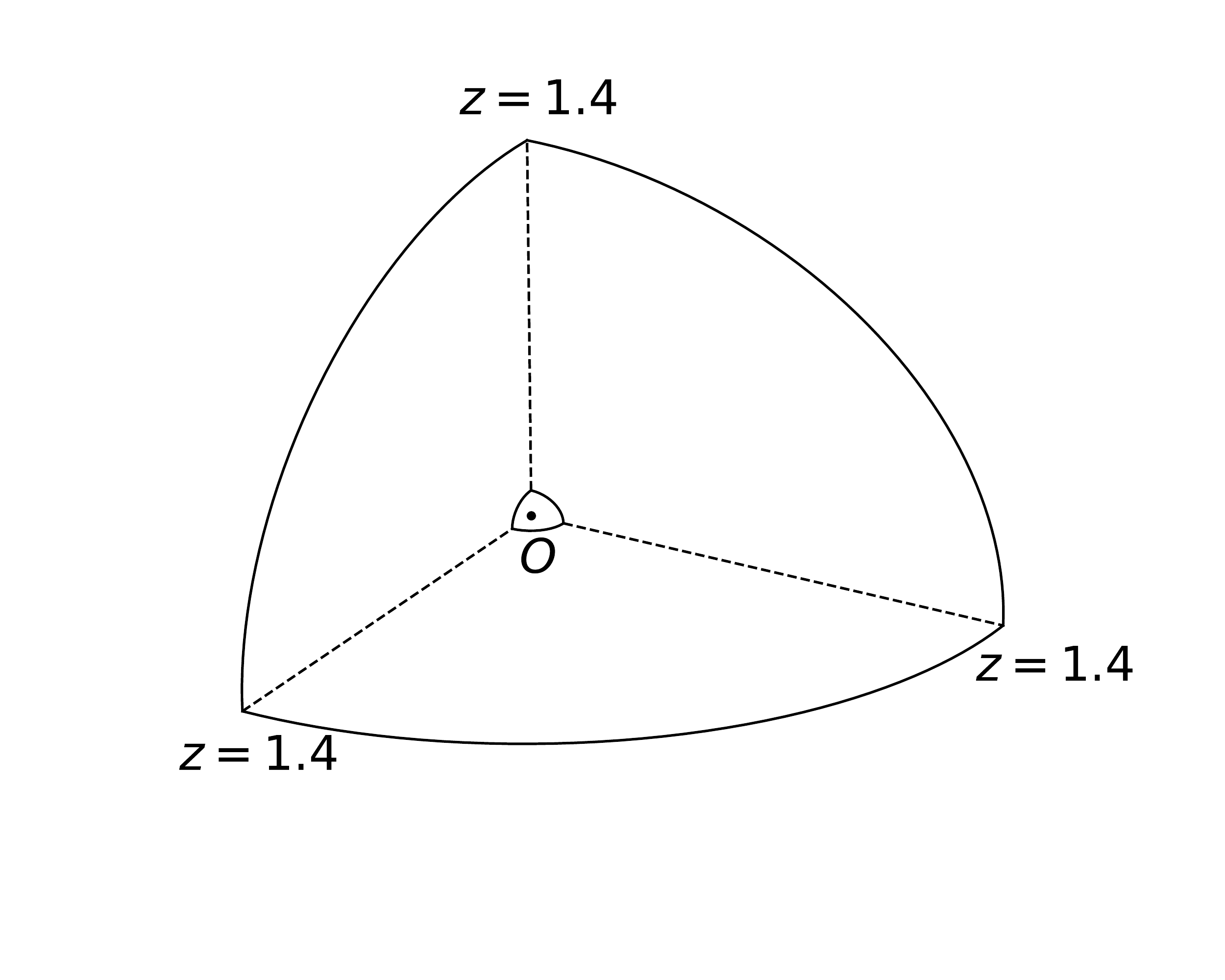}{The survey region of the MICE-GC simulation, which is an octant. Note that due to some technical reasons there are no galaxies near the origin $O$, so a small octant is removed from the survey region.}

We consider the halo catalog in this light cone with halo masses between $2.2\times 10^{12}h^{-1}M_{\odot}<M<10^{14}h^{-1}M_{\odot}$, a wide mass bin. We obtained similar results using other mass bins as well. The effective redshift of this light cone is $z_{\rm eff}=0.76$. We assume the bias parameter $b_{1}$ and $b_{2}$ to be free parameters of the model, and use FAST-PT \cite{McEwen:2016fjn} to determine the bias parameters to be:
\bea 
b_1&=&1.88\vs
b_2&=&3.13\,.
\eea
The remaining bias parameter $b_{s^2}$ can be constrained by assuming the bias model is local in Lagrangian space \cite{Baldauf:2012hs}:
\be 
b_{s^2}=-\frac{4}{7}(b_1-1)=-0.50\,.
\ee 
We use the quadratic estimator, \ec{qeLC} to obtain the reconstructed Fourier space windowed galaxy density field, and transform it back into real space. Then, we compare this indirectly estimated result with the directly measured (using the FKP estimator) galaxy density in real space in Fig.~\rf{real_fkp}. We use the information from small scale modes up to $k_{s,\rm max}=0.48\,h\,\rm Mpc^{-1}$. We also plot in Fig.~\rf{SN} the directly measured large scale galaxy power spectrum with cosmic variance error bars (corresponding to \ec{errorLC} with $\mathpzc{A}=0$) versus the estimated large scale power spectrum using \ec{qeLC} with detectability given by \ec{errorLC}. We see that our quadratic estimator gives a good estimation of the linear matter power spectrum and the uncertainty of the estimated
result is only slightly larger than cosmic variance.

Note that in Fig.~\rf{real_fkp} because of the light cone, we
cannot have a direct measurement of the
$\delta_{g,W}$ field.  Both the field computed
from direct measurement of the Fourier modes (with FKP weighting) and
the field derived from the quadratic estimator
can be seen to encode almost the same large
 scale information from the light cone catalog. The large scales we
 are observing correspond to about $10^{-3}h\,\rm Mpc^{-1}$$<k_l<10^{-2}h\,\rm
 Mpc^{-1}$, where the magnitude of the observed power spectrum
 Fig.~\rf{SN} is much greater than the shot noise term  (in this
 case $P_{\rm shot \; noise}\simeq 1000\,(h^{-1}\rm Mpc)^3$). From
 Fig.~\rf{real_fkp} we see that as our quadratic estimator  extracts
 large scale information, the cells with  large over- and
 under-densities are especially well reconstructed. The difference
 with the true field becomes
 slightly larger when we go to higher redshift (corresponding to the
 panels on the right) and is worst for the very right panel.
 
 \Sfig{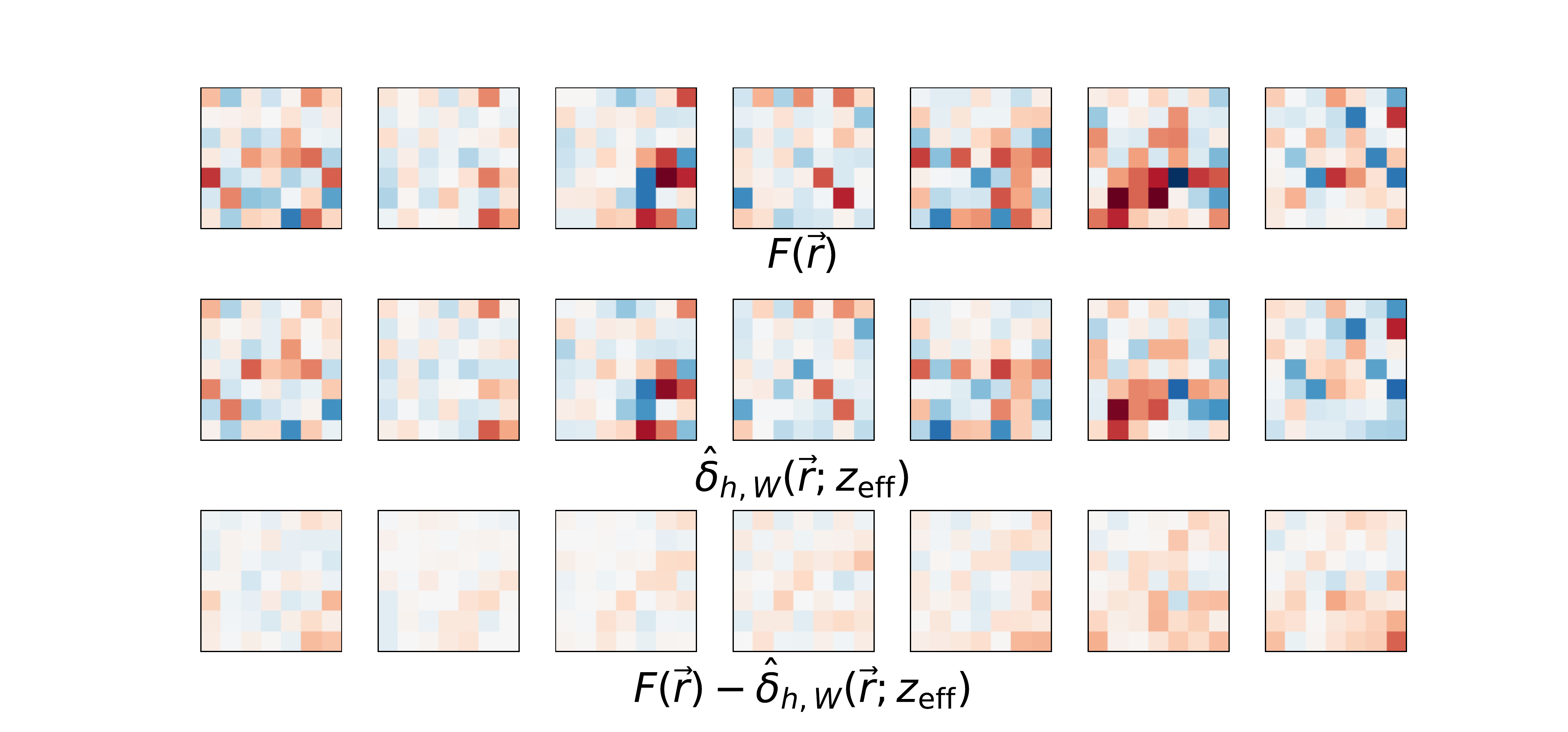}{Comparison of the true real space  galaxy density field in the MICE-GC simulation ($F(\vec{r})$ computed using the directly measured large scale modes and FKP estimator, top row) and the windowed halo density field from the quadratic estimator ($\hat{\delta}_{h,W}(\vec{r})$, middle row). The bottom row shows their difference. Each cell is ($0.44\,h^{-1}\rm Gpc$) thick,
   and panels are arranged so that the mean redshift increases from
   left to right.
   The upper limit on $\vec{k}_s$ input to the quadratic estimator is $0.48\,h\,\rm Mpc^{-1}$.}

\Sfig{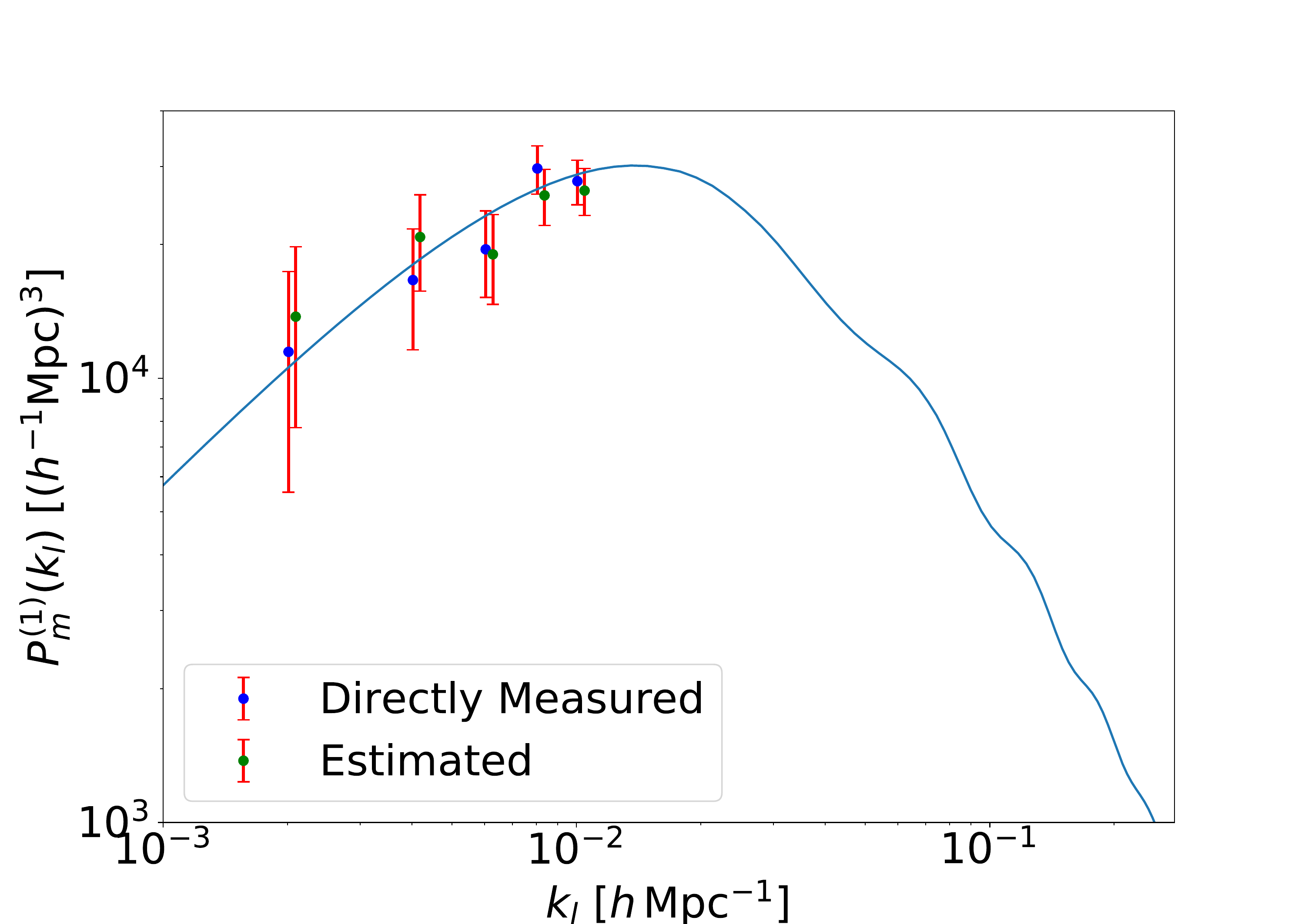}{The inferred linear matter power spectrum from direct measurement versus indirect estimation using our quadratic estimator, both from the MICE-GC light-cone halo catalog. The error bar for the direct measurement is derived from  cosmic variance (see text). The error for the indirect estimate is expressed using  $P_{m}^{(1)}(k_l)\sigma(k_l)$, where $\sigma(k_l)$ is from \ec{errorLC} after scaling.}

\section{Conclusion} \label{sec5}
\subsection{Summary}
In prior work \cite{Li:2020fir} we have shown that the amplitude and phase of large scale density fluctuations can be recovered by applying a quadratic estimator to measurements of small scale Fourier modes and their correlations. In this paper we extend that work (which was limited to a matter density field at a single instance in cosmic time) to a light cone galaxy catalog in order to make it applicable to observational data. All extensions are tested on appropriate mock survey datasets derived from N-body simulations.\

\subsection{Discussion}

Our formalism includes the major effects that are relevant for an application to observational data. There are some minor aspects however which will need to be dealt with when this occurs. One is the fact that we have tested on homogeneous mock surveys, when real observations will include masked data (to account for bright stars for example), and a potentially more complex window function. In a spectroscopic survey, the observed distribution of galaxies is distorted and squashed when we use their redshift as an indicator of their radial distance due to galaxies' peculiar velocity. This effect is known as redshift space distortion \cite{Kaiser:1987rsd} and the FKP formalism corresponds to the monopole moment in redshift space. We have left the generalization to include higher order power spectrum multipoles (quadrapole, hexadecapole) to future work.

At present the large scale limitations on direct measurement of galaxy clustering are observational systematics (e.g., \cite{Ho:2012sh}). These include angular variations in obscuration, seeing, sky brightness, colors, extinction and magnitude errors. Because these result in relatively small modulations of the measured galaxy density, they will affect large scale modes most importantly, hence the utility of our indirect measurements of clustering on these scales. Quantification of these effects on the scales for which we do measure clustering will still be needed though. It will be also be instructive to apply large scale low amplitude modulations to our mock surveys in order to test how well the quadratic estimator works with imperfect data. Even small scale issues with clustering, such as fiber collisions \cite{Hahn:2016kiy} could affect our reconstruction, depending on how their effects propagate through the quadratic estimator.

Observational datasets exist at present which could be used to carry out measurements using our methods. These include the SDSS surveys BOSS \cite{Dawson:2013boss} and eBOSS \cite{Dawson:2015wdb} (both luminous red galaxies and emission line galaxies). Substantial extent in both angular coordinates and redshift are necessary, so that deep but narrow surveys such as VIMOS \cite{Fevre:2014tna} or DEEP2 \cite{Coil:2005ap} would not be suitable. In the near future, the available useful data will increase rapidly with the advent of WEAVE \cite{Dalton:2014wv} and DESI \cite{DESI:2019ds}. Space based redshift surveys  with EUCLID \cite{Amiaux:2012ec} and WFIRST \cite{Wfirst:2012jg} will expand the redshift range, and SPHEREx \cite{Dore:2014cca}, due for launch even earlier will offer maximum sky coverage, and likely the largest volume of all.

In order to model what is expected from all these datasets, the effective range of wavelengths used in the reconstruction of large scale modes should be considered. Surveys covering large volumes but with low galaxy number density will have large shot noise contributions to density fluctuations, and this will limit the range of scales that can be used. For example, in our present work we have successfully tested number densities of $\sim 3\times10^{-3}$ galaxies per (Mpc/h)$^{3}$. Surveys such as the eBOSS quasar redshift survey \cite{Ata:2017dya} with a number density $\sim 100$ times lower will not be useful, for example.

Once an indirect measurement of large scale modes has been made from an observational dataset, there are many different potential applications. We can break these up into two groups, involving the power spectrum itself, and the map (and statistics beyond $P_{\rm m}(k)$) which can be derived from it.

First, because of the effect of observational systematics mentioned above, and the fact the our indirect estimate of clustering is sensitive to fluctuations beyond the survey boundaries itself, then it is likely that the measurement we propose would correspond to the largest scale estimate of three dimensional matter clustering yet made. This would in itself be an exciting test of theories, for example probing the power spectrum beyond the matter-radiation equality turnover, and allowing access to the Harrison-Zeldovich portion. There has been much work analyzing large scale anomalies in the clustering measured from the CMB \cite{Copi:2010na}\cite{Rassat:2014yna}\cite{Schwarz:2015cma}, and it would be extremely useful to see if anything comparable is seen from galaxy large scale structure data. On smaller scales, one could use the matter-radiation equality turnover as a cosmic ruler \cite{Hasenkamp:2012ii}, and this would allow comparison to measurements based on BAO \cite{Lazkoz:2007cc}.

Second, there will be much information in the reconstructed maps of the large scale densities (such as Fig.~\rf{real_fkp}). One could look at statistics beyond the power spectrum, such as counts-in-cells \cite{Yang:2011cic}, or the bispectrum, and see how consistent they are with model expectations. One can also compare to the directly measured density field and obtain information on the large scale systematic effects which are modulating the latter. Cross-correlation of the maps with those of different tracers can also be carried out. For example the large scale potential field inferred can be used in conjunction with CMB observations to constrain the Integrated Sachs Wolfe effect\cite{Nishizawa:2014vga}.

In general, as we will be looking at large scale fluctuations beyond current limits by perhaps an order of magnitude in scale or more, one may expect to find interesting constraints on new physics. For example evidence for the $\Lambda$CDM model was seen in the first reliable measurements of large scale galaxy clustering on scales greater than $10\,h^{-1}\rm Mpc$ (e.g., \cite{Efstathio:1990cdm}). Moving to wavelengths beyond $2\pi/(k=0.02) \sim 300 \rm \, Mpc$ may yet lead to more surprises.

\acknowledgements
We thank Jonathan A. Blazek, Duncan Campbell and Rachel Mandelbaum for resourceful discussions. We also thank Enrique Gaztanaga for providing us with the $1$ in $700$ matter particles' positions of Mice-GC simulation for a test. This work is supported by U.S. Dept. of Energy contract DE-SC0019248 and NSF AST-1909193.
This work has made use of CosmoHub. CosmoHub has been developed by the Port d'Informació Científica (PIC), maintained through a collaboration of the Institut de Física d'Altes Energies (IFAE) and the Centro de Investigaciones Energéticas, Medioambientales y Tecnológicas (CIEMAT), and was partially funded by the ``Plan Estatal de Investigación Científica y Técnica y de Innovación" program of the Spanish government.
\clearpage

\end{document}